\documentstyle[prl,aps,twocolumn,epsf]{revtex}
\large  

\begin{document}
\draft \twocolumn[\hsize\textwidth
\columnwidth\hsize\csname
@twocolumnfalse\endcsname

\title{Sign-time distributions for interface growth} 
\author{Z. Toroczkai$^{(1,2)},
$ T. J. Newman$^{(1)}$, 
and S. Das Sarma$^{(2)}$} 
\address{$^{(1)}$Department of Physics, Virginia 
Polytechnic Institute and State University, Blacksburg 
VA 24061\\ $^{(2)}$Department of Physics,
University of Maryland, College Park, MD 20742} 
\maketitle
\begin{abstract}
We apply the recently introduced distribution of sign-times 
(DST) to non-equilibrium interface growth dynamics. 
We are able to treat within a  unified 
picture the persistence properties of a large class of 
relaxational and noisy linear growth processes, and prove the 
existence of a non-trivial scaling relation. 
A new critical dimension is found, relating to the 
persistence properties of these systems. 
We also illustrate, by means of numerical simulations, 
the different types of DST to be expected in both linear 
and non-linear growth mechanisms. 
\end{abstract}
\vspace{5mm} \pacs{PACS numbers: 05.40.+j, 05.70.Ln, 82.20.Mj } ]

The notion of persistence, or the statistics of first passage 
events, has been a powerful conceptual tool in studying
stochastic non-Markovian processes in many research areas of
physics, engineering, statistics, and applied mathematics.
In this Letter we apply a persistence-related concept, the 
distribution of sign-times, or DST, (defined below) 
to the problem of kinetic surface roughening in 
non-equilibrium interface growth dynamics \cite{krug}. 
We believe that the ideas,
described in this paper, could become
an extremely useful conceptual and practical tool
in characterizing surface growth dynamics, rivaling 
the dynamic scaling ideas currently used in studying 
kinetic surface roughening. Depending on the specific
issues of interest, our proposed DST technique may 
actually be more powerful and informative than the
currently fashionable dynamical/roughness/growth 
exponent based characterization of dynamical surface
morphologies.

One of the main themes in the theory of non-equilibrium
interfaces is grouping the interface roughening phenomena
within `universality classes'. This classification of
scenarios is based on calculating the dynamic scaling 
properties of the surface correlation function \cite{krug}.
On the other hand, in  non-equilibrium interface growth 
experiments, 
one might also be interested in 
morphology stability issues which can in fact
be formulated as first passage type questions:
what is the probability
that a mound (or crevice) will survive as a mound
(crevice) for a given period of time $t$? How does
this probability decay in time, etc.?
These type of questions however are not simply 
delineated by such a correlation function. 

Another open theoretical problem is to establish
a correspondence between discrete
solid-on-solid (SOS) models
and continuum Langevin equations beyond the 
equality of exponents. 
For example,
based on structure factor measurements, the authors
in Ref. \cite{KD} claim that the SOS model they 
introduced does not only belong to the same universality 
class as the noisy Mullins equation but it is 
described {\em exactly} by it. Our approach proposed 
in the present Letter (which is {\em not} based on direct
measurement of 
the correlation function)  supports that claim. 

It would be useful therefore, to study statistical quantities
that are directly sensitive to the structural and morphological 
properties of interfaces (e.g., formation of mounds) and  to 
the dynamics of these structures (e.g., coarsening).  
In this Letter, we propose that such information may be
inferred from the DST, which
has recently been introduced in the context of 
the persistence properties of simple coarsening systems 
and the diffusion equation\cite{dg,nt}. 
First passage time or persistence problems have been
the focus of intensive research for the past few years, 
producing a series of analytic and
numerical results with applications to the Ising and Potts 
models\cite{potts}, the diffusion equation \cite{de},
phase ordering \cite{po}, interface kinetics \cite{ik}, 
etc. and experiments 
on liquid crystals and soap froths 
(see references in \cite{de}).
The central issue of persistence concerns the probability
of an event {\em never} occurring (up to a certain time $t$).
It is very restrictive by definition, 
and  good statistics from numerics or experiments may 
be extremely hard to obtain. The recently introduced
\cite{dg,nt} DST is practically more 
accessible, and as a limiting case produces the 
persistence probability. 

The DST is essentially a histogram
performed on the sign of the fluctuations 
and simply measures the probability of the 
fluctuations having been in the positive domain for a 
total time $\tau$ in the
given time $t$ of the process. 
Obviously for $\tau=t$ we obtain
the usual persistence probability, 
which we denote by $P_{+}(t)$, and for $\tau=0$ we
obtain the probability of the
fluctuations having {\it never} been in the positive domain, 
i.e, to have been
{\em always} in the negative domain, $P_{-}(t)$. 
The distinction between the persistence
of fluctuations in the positive domain and in the 
negative domain becomes important in the case of
nonlinear models \cite{kk}. We shall refer to these
as `positive' and `negative' persistence, respectively. 

The sign-time for an interface on a $d$-dimensional 
substrate is the stochastic variable defined by:
\begin{equation}
T({\bf x},t) = \int \limits _{0}^{t} dt' H
\left( h({\bf x},t) \right )  \ , 
\label{st}
\end{equation} 
where $H$ is the Heaviside step function and 
$h({\bf x},t)$ is the height of the
interface  measured with respect to the average height. 
Since $h$ is a random variable (due to its coupling 
to the noise) the sign-time will 
be described by a probability
distribution -- the DST. For a system with translation 
invariance, the statistics of $\tau $ will not depend on the
location 
${\bf x}$, and so the DST may be written
as
\begin{equation}
\label{std}
S(\tau,t) = \langle \delta 
(\tau -T({\bf 0},t)) \rangle \ ,
\end{equation}
where $\langle .\rangle$ indicates the average 
over the noise.  Some properties of $S$ are: 
i) it is defined on 
the interval $0 \le \tau/t \le 1$; ii) for interface
growth with $h \rightarrow -h$ symmetry, $S$ will be 
symmetric about $\tau/t=1/2$; iii) the tails of the 
distribution give the persistence probabilities:
$P_+(t)=\int_{0}^{\epsilon} d\tau\;S(\tau,t)$ and
$P_-(t)=\int_{0}^{\epsilon} d\tau\;S(t-\tau,t)$,
where $\epsilon \ll t$ is a microscopic time scale (of the 
order of the 
fastest temporal scale in the interface dynamics
). These  probabilities are expected
to have a power law decay, defining the corresponding 
persistence exponents $\theta_{\pm}$:
$P_{\pm}(t) \sim (1/t)
(\epsilon/t )^{\theta _{\pm}-1}$,  iv)
the shape of $S$ contains 
information about whether the growth is rare event 
dominated or not. 

In the spirit
of ref.\cite{ik} we first consider the 
following class of stochastic 
linear equations:
\begin{equation}
\partial _{t} h = 
-\nu (-\nabla ^{2})^{z/2}h + \xi \ ,
\label{langevineq}
\end{equation}
with flat ($h({\bf x},0)=0$) initial condition,
 where $\xi$ is a general noise 
term which may represent
the `pure deterministic' case via the 
choice $\xi({\bf x},t)=
\delta(t)\eta({\bf x},t)$ or 
the regular `noisy' case with
$\xi({\bf x},t)=\eta({\bf x},t)$
where $\eta$ is a 
Gaussian-distributed noise possibly with 
spatial correlations. We 
consider the following three 
choices for $\eta$: 1) white
noise with correlator $\langle 
\eta({\bf x},t)
\eta({\bf x'},t')\rangle=2D\delta({\bf x}
-{\bf x'})\delta(t-t')$,
2) volume conserving noise $\langle
 \eta({\bf x},t)
\eta({\bf x'},t')\rangle=-2D 
\nabla^2 \delta({\bf x}-
{\bf x'})\delta(t-t')$ and 
3) long range spatially
correlated noise $\langle 
\eta({\bf x},t)
\eta({\bf x'},t')\rangle=2D |{\bf x}-
{\bf x'}|^{\rho-d}\delta(t-t')$,
$\rho < d$. For example, the Edwards-Wilkinson (EW)
model may be recovered by setting $z=2$ in 
Eq. \ref{langevineq}, and by applying white noise;
likewise the noisy Mullins equation corresponds to
setting $z=4$ \cite{krug}. We write Eq.(\ref{std}) through the
higher moments of DST as
\begin{equation}
S^{(z)}_d(\tau,t) = \sum_{n=0}^{\infty}
\int\limits _{-\infty}^{\infty}
{d\omega \over 2\pi }
e^{i\omega \tau} \frac{(-i\omega)^{n}}{n!} 
\langle 
\ [T^{(z)}_d({\bf 0},t)]^n \rangle \ ,
\label{stde}
\end{equation}
where we have introduced a frequency representation
of the delta function, and expanded in powers of the
sign-time $T^{z}_{d}$.
We shall enter into no technical 
details here on how to proceed with
calculating the moments of the DST.
We present only the final form
that we obtained for the $n^{\rm th}$ order 
moment normalized by $t^{n}$
($ \mu_n \equiv\langle 
(\tau/t)^n \rangle$):
\begin{eqnarray}
\mu_n=\prod_{k=1}^{n}\int\limits_0^{1} 
\frac{da_k}{2\pi}\int\limits_{-\infty}^{\infty}
\frac{d\sigma_k}{\epsilon_k+i\sigma_k}\; 
e^{-\sum_{j,l}\sigma_j
\sigma_l \kappa(a_j,a_l)} \ ,  
\label{nmom}
\end{eqnarray} 
where the limits of $\epsilon_k \to 0^+$ are to be taken, and
\begin{eqnarray}
\kappa(x,y) = \left \lbrace 
\begin{array}{l}(x+y)^{-\gamma},\;\;
\mbox{deterministic case} \\
\int\limits_{0}^{min(x,y)} du (x+y-2u)^{-\gamma},
\;\; \mbox{noisy case}
\end{array} \right. \label{kappa}
\end{eqnarray} 
with $0\leq x,y \leq 1$, and $\gamma$ being given by 
1) $\gamma=d/z$ for the deterministic case and for white noise, 
2) $\gamma=(d+2)/z$ for volume conserving noise, and 
3) $\gamma=(d-\rho)/z$ for long-range correlated noise.
We make the following observations from 
Eqs. (\ref{stde})-(\ref{kappa}). First, the DST obeys the exact 
scaling form
\begin{equation}
S^{(z)}_d(\tau,t) = \frac{1}{t}F_{\gamma}
\left( \frac{\tau}{t} \right),
\;\;0 \leq \tau\leq t \ ,
\label{scale}
\end{equation}
for {\em all} values of $t$ ($\mu_n$ is $t$-independent). 
Second, the `material parameters' $\nu $ and 
$D$ do not appear in the DST.
Third, the three numbers $(d,z,\rho )$
appear in the DST (for any 
$t$) {\em only} through their
combination in $\gamma=\gamma(d,z,\rho)$.
Thus, the persistence exponents (which are contained
within the DST) will also only depend on 
$d$, $z$ and $\rho$ through the exponent $\gamma $.
[This appears to be implicitly understood in ref.\cite{ik}
where persistence is measured as a function of 
the growth exponent $\beta = {\rm max}(0,(1-\gamma)/2)$].
A similar scaling property for the persistence exponents is also true 
for the deterministic case. 

\begin{figure}[htbp]
\begin{minipage}{1.7 in}\epsfxsize=1.6 in 
\epsfbox{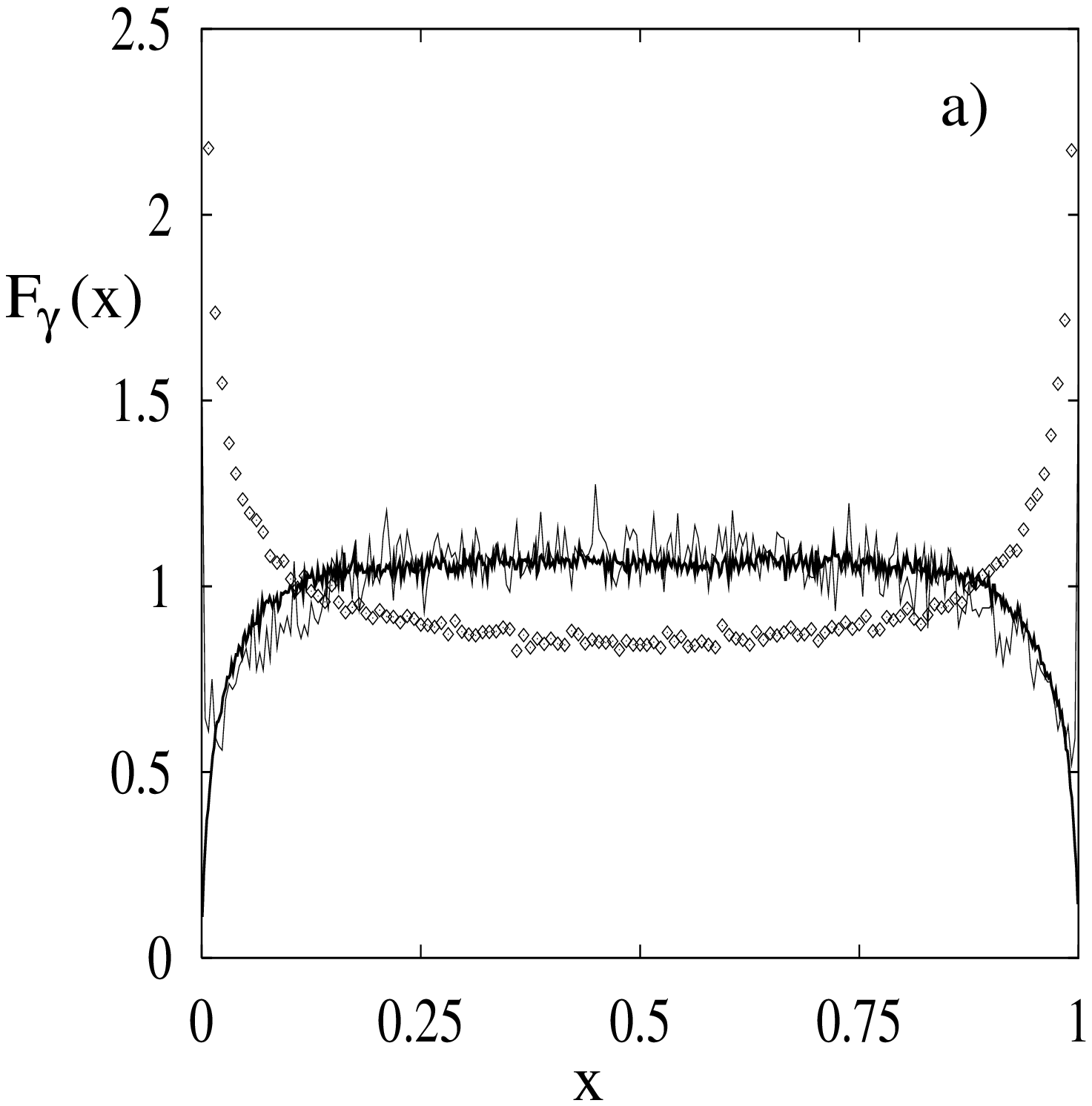}
\end{minipage}
\hspace*{-0.25cm}
\begin{minipage}{1.7 in}\epsfxsize=1.6 in 
\epsfbox{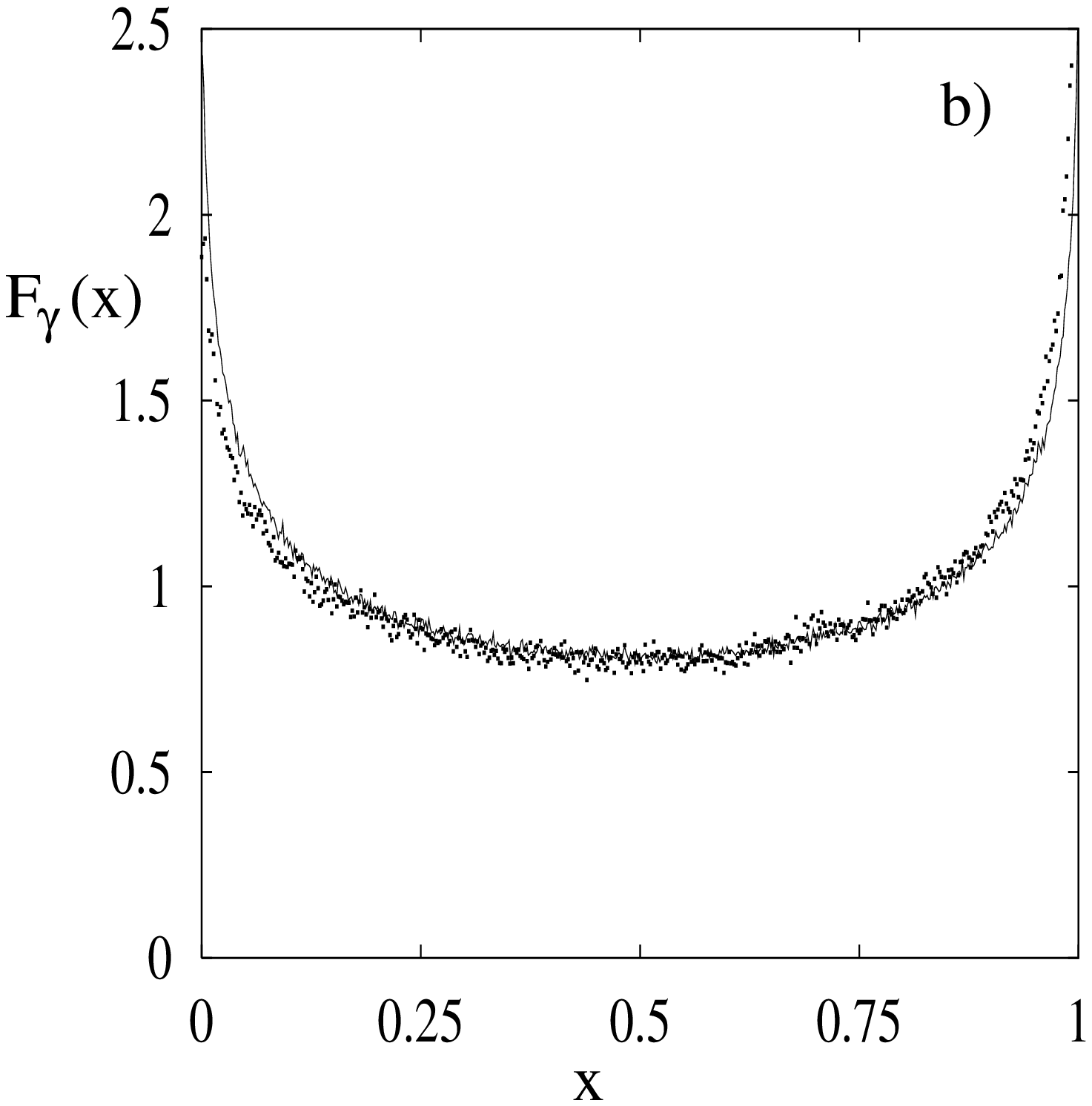}
\end{minipage}
\vspace*{0.5cm}
\caption{DST's ($F_{\gamma}(x)=t S(xt,t)$) 
for a) $\lbrace 1,2 \rbrace$ 
(thick line) at 
$t=0.25\times 2\cdot 10^3$ obtained on a 
grid of $L=2048$ sites and averaged over $2\cdot 
10^3$ runs; $\lbrace 2,4 \rbrace$
at $t=0.01\times 256$ (diamonds), 
and $t=0.01\times 4096$ (thin line), 
on a grid of $1024\times 1024$
shown for a single run,
b) the SOS large curvature model (dots) on a lattice 
of $L=10^4$ at $8192$ steps, averaged over 100 runs; and for 
$\lbrace 1,4 \rbrace$ measured 
on a grid of $L=2048$ 
sites at $t=0.05\times 2\cdot 10^3$, and averaged over
$2\cdot 10^3$ runs.}\label{Compare}
\end{figure}

For simplicity of the
notation, instead of $S^{(z)}_d(\tau,t)$ 
(and $\theta^{(z)}_d$) we will use $S_{\gamma}(\tau,t)$
(and $\theta_{\gamma}$). 
Let us consider as an example 
the generic case of white noise, for which
$\gamma=d/z$. According 
to the above, for any 
model for which, e.g., $d/z=0.5$, the DST 
(and thus the persistence
properties) will be identical to 
that for the EW model in one
dimension. We compared the numerically
obtained DST's for $\lbrace 1,2 \rbrace$ (meaning $d=1$, $z=2$) 
and $\lbrace 2,4 \rbrace$ . According to the above, one should observe identical DST's. 
The $\lbrace 1,2 \rbrace$ DST was measured using a standard 
discretization scheme, see Fig.1a). 
The numerical integration of the case $\lbrace 2,4 \rbrace$
is less straightforward. We used the simplest
discrete scheme for modeling
the operator $\nabla ^{4}$ in $d=2$, as more sophisticated
schemes were actually less stable under the influence
of additive white noise. A very small integration time step
of $\delta t = 0.01$ was used to ensure stability. We observed
a long transient in which the DST was actually concave,
in contrast to the convex DST for the case $\lbrace 1,2 \rbrace$
($d=1$ EW model). 
After $10^3$ iterations the DST began to turn over, and
eventually settled into a convex shape, closely matching
the $\lbrace 1,2 \rbrace$ DST. This illustrates the
sensitivity of the DST to lattice effects, which may be a very
useful property if one is investigating physics which is
itself sensitive to the underlying lattice. In fig.1b) we
show the DST's obtained from numerical integration of the
case $\lbrace 1,4 \rbrace$ (the $d=1$ noisy Mullins equation)
and numerical simulation of the SOS large 
curvature model\cite{KD}.
These two models are expected to have very similar 
properties, and indeed their DST's are almost indistinguishable.

It follows from Eqs. (\ref{stde})-(\ref{kappa}) that the 
knowledge of the second moment $\mu_2$ {\em uniquely} 
determines the value of $\gamma$,
and therefore also $z$ (in a given dimensionality).
It is possible to evaluate
analytically the second moment $\mu_2$. We find 
$\mu_2=1/2 - G(\gamma)$, where for the deterministic case
\begin{equation}
G(\gamma)= \frac{\gamma}{4\pi}
\int\limits_0^1 da\;\frac{1-a}{1+a}\;\left[
\left(\frac{1+a}{2a^{1/2}} \right)^{2\gamma}-1
\right]^{-1/2} \ ,
\label{det2}
\end{equation} 
and for the noisy case:
\begin{equation}
G(\gamma)= \frac{1}{2\pi}
\int\limits_0^1 da\;\mbox{arctg}\sqrt{
\frac{(4a)^{1-\gamma}}
{\left[(1+a)^{1-\gamma}-(1-a)^{1-\gamma}
 \right]^{2}} -1} \ .
\label{nos2}
\end{equation} 
The second moment in both cases is a monotonic function
of $\gamma$ and therefore the knowledge of one determines
the other; a property useful in deciding whether a 
measured DST can indeed be described by a process 
like (\ref{langevineq}). For example, one may obtain from 
numerical or experimental measurements a symmetric DST, 
from which one may compute $\mu_2$. One can {\em test} 
therefore if the process generating the measured DST can be 
described by Eq. (\ref{langevineq}): one determines  
$z$ using the above procedure, and then simulates Eq. 
(\ref{langevineq}) with the corresponding value of $\gamma$, 
thus generating a new DST. If the two DST's are very close or 
coincide then the assumption that the physical process may be
modeled by Eq.(\ref{langevineq}) is valid, just as in 
the SOS large curvature model case, shown in Fig.1b). 
Note, that this procedure also requires an assumption about
the type of noise.  
 
The integral in (\ref{kappa}) 
is divergent for $\gamma > 1$ at $x=y$.
Introducing a microscopic lattice cut-off, the DST can  be 
calculated \cite{big} to give a Dirac $\delta$-function 
centered around $\tau=t/2$:
\begin{equation}
S_{\gamma}(\tau,t)=\frac{1}{t} \delta \left(\frac{1}{2} 
-\frac{\tau}{t} \right)\;,\;\;\;\mbox{for any}\;\;\;
\gamma>1 \ .
\label{delta}
\end{equation} 
It is a well known result \cite{krug}
that for Eq. (\ref{langevineq}), $d=d_u=z$ is an upper
critical dimension  and separates interfaces which are
asymptotically rough from those which are asymptotically
smooth. Thus, the fact that for dimensions above $z$ there 
is no roughening, is reflected by a $\delta$-function DST, 
i.e., {\em all} points of the interface will spend 
exactly half of their
time above the mean height. The persistence exponent in 
this case is not really defined, since the persistence 
probability is zero. 
In a lattice model, one would expect corrections to scaling 
to the above result, and for the persistence 
probability to decay exponentially with time. 
When approaching $\gamma_u \equiv d_u/z=1$ from below, 
the persistence exponent
diverges; $\gamma=\gamma_u$ being a marginal case for which 
no numeric or analytic results have been produced yet (on 
persistence properties). It is precisely the case of the two 
dimensional EW equation with white noise. Note that 
the EW equation in any integer dimension 
($d \geq 1$) with volume conserving 
noise is in the smooth phase, and would therefore have 
a $\delta $-function DST ($\gamma=1+d/2$), with the 
persistence exponent undefined (or formally infinite).

\begin{figure}[htbp]
\epsfxsize = 3.4 in \epsfbox{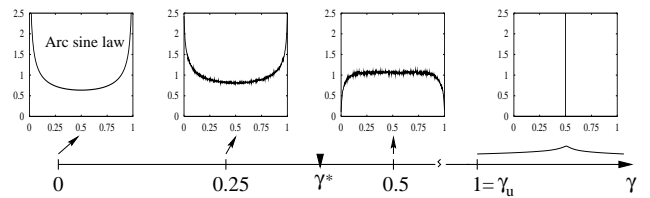}
\vspace*{0.5cm}
\caption{The behavior of DST as a function of $\gamma$ 
for eq. (\ref{langevineq}),  
with white noise. Each inset shows the function
$F_{\gamma}(x)$ vs. $x$. 
For $\gamma=0.25$ and 
$\gamma=0.5$ we simulated the Langevin equation
with white noise in one dimension at $z=4$ and 
$z=2$, respectively. For $z=2$ the simulation parameters
were the same as for the thick line in fig.1a) 
and for $z=4$ they were the same as for 
the continuous line in fig.1b) }\label{map}
\end{figure} 

From the scaling relation (\ref{scale}) one can infer the
existence of a new critical `dimension' $\gamma^{*}$ both
for the deterministic and noisy cases: since the
tails of DST give the persistence probability,
which has a power law  decay ($P_{\pm}\sim t^{-\theta}$), 
the scaling function $F_{\gamma}$ must obey the behavior 
$F_{\gamma}(x)\sim x^{\theta-1}$, for $x \ll 1$, and  
$1-x \ll 1$ in order that (\ref{scale}) be satisfied.
For $\theta < 1$ the DST has integrably divergent tails
while for $\theta > 1$ the tails vanish (as $x^{\theta-1}$).
In the former case the sites are more likely to be found in 
a positive or negative persistent state
(i.e. with a height that did
not change sign at all) , while in the latter case
persistent sites will be an absolute minority 
(with vanishing measure as $t\to \infty$). 
Since $\theta_{\gamma}$ is a monotonically increasing
function of $\gamma$ the equation $\theta_{\gamma}=1$ will be
satisfied at a unique value of $\gamma^*$. At this value
$F_{\gamma^*}$ is flat at the tails, 
it neither falls to zero nor diverges.
This shows that the value of $\theta_{\gamma^*}
\equiv \theta^{(z)}_{d^*} = 1$ is special. 
It is possible that $F_{\gamma^*}$ can still have 
some structure around $\tau/t = 1/2$, but the 
simplest possibility is that it is a top-hat function. In this
case $\mu_2=1/3$,
and therefore $\gamma^*$ can be calculated
after (numerically) inverting $G(\gamma)=1/6$, using Eqs.
(\ref{det2}) and (\ref{nos2}). For the deterministic case
one obtains $\gamma^*=17.983..$ and for the noisy case
$\gamma^*=0.438..$. 
For the noisy case, our numerical simulations
are compatible with $0.25 < \gamma ^{*} < 0.5$, as can be 
seen from Fig.2. It is interesting to note that the
permanent presence of noise `brings down'
this critical $\gamma^*$ to a sub-unitary value as compared
to the deterministic case. 

The DST for $\gamma = 0$ is exactly 
known, and is called the `arcsine law' in the mathematical
literature\cite{levy}: 
$F_{0}(x)=1/(\pi\sqrt{x(1-x)})$, which can also be
derived from Eqs. (\ref{stde})-(\ref{kappa}) presenting  a 
novel alternative to 
this venerable old problem.
Fig.2 summarizes our findings on the different regimes for 
the DST of the noisy case
of Eq. (\ref{langevineq}) with the two critical
`dimensionalities' $\gamma^*$ and $\gamma_u$. 

\begin{figure}[htbp]
\begin{minipage}{1.7 in}\epsfxsize=1.6 in 
\epsfbox{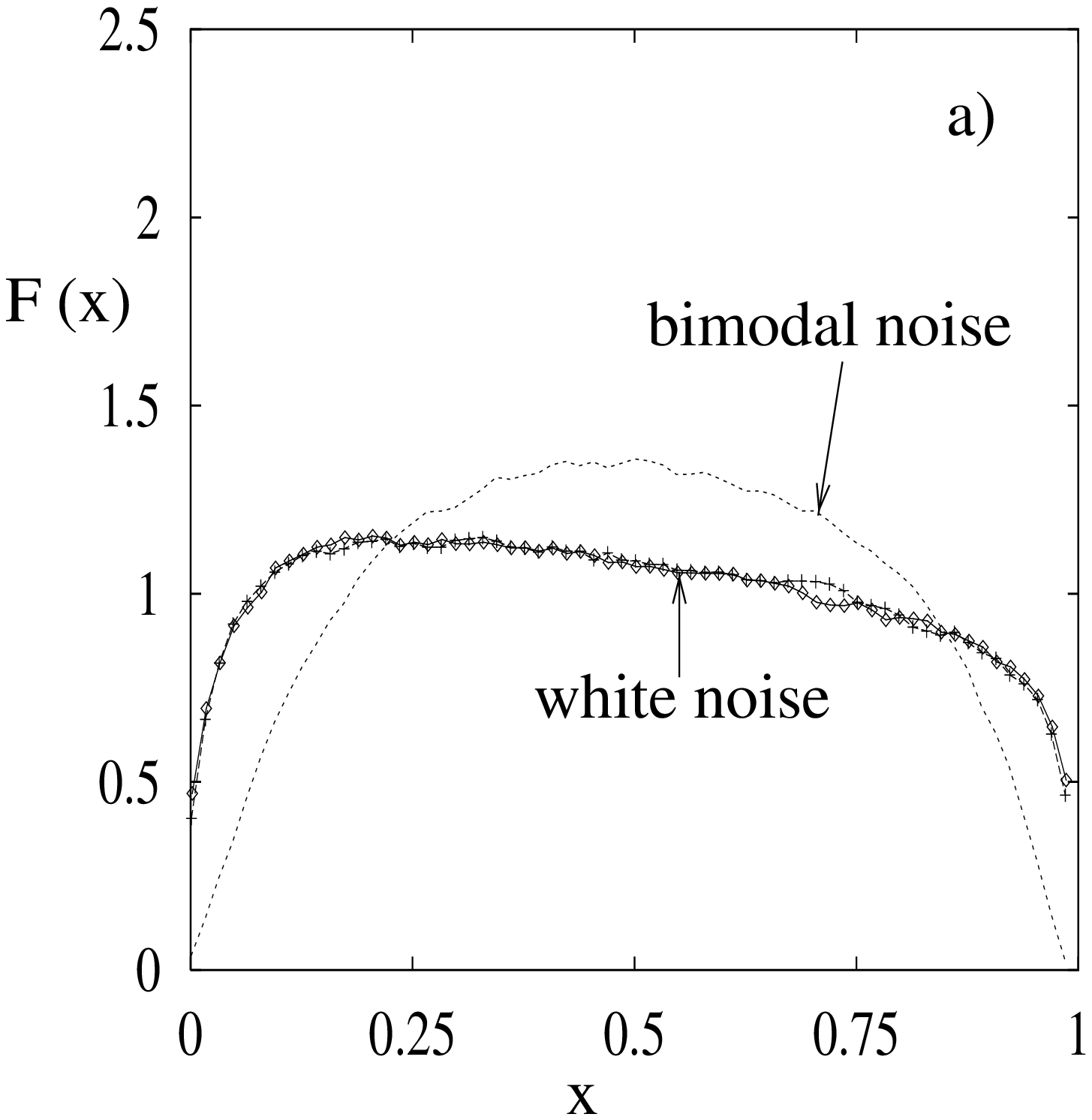}
\end{minipage}
\hspace*{-0.25cm}
\begin{minipage}{1.7 in}\epsfxsize=1.6 
in \epsfbox{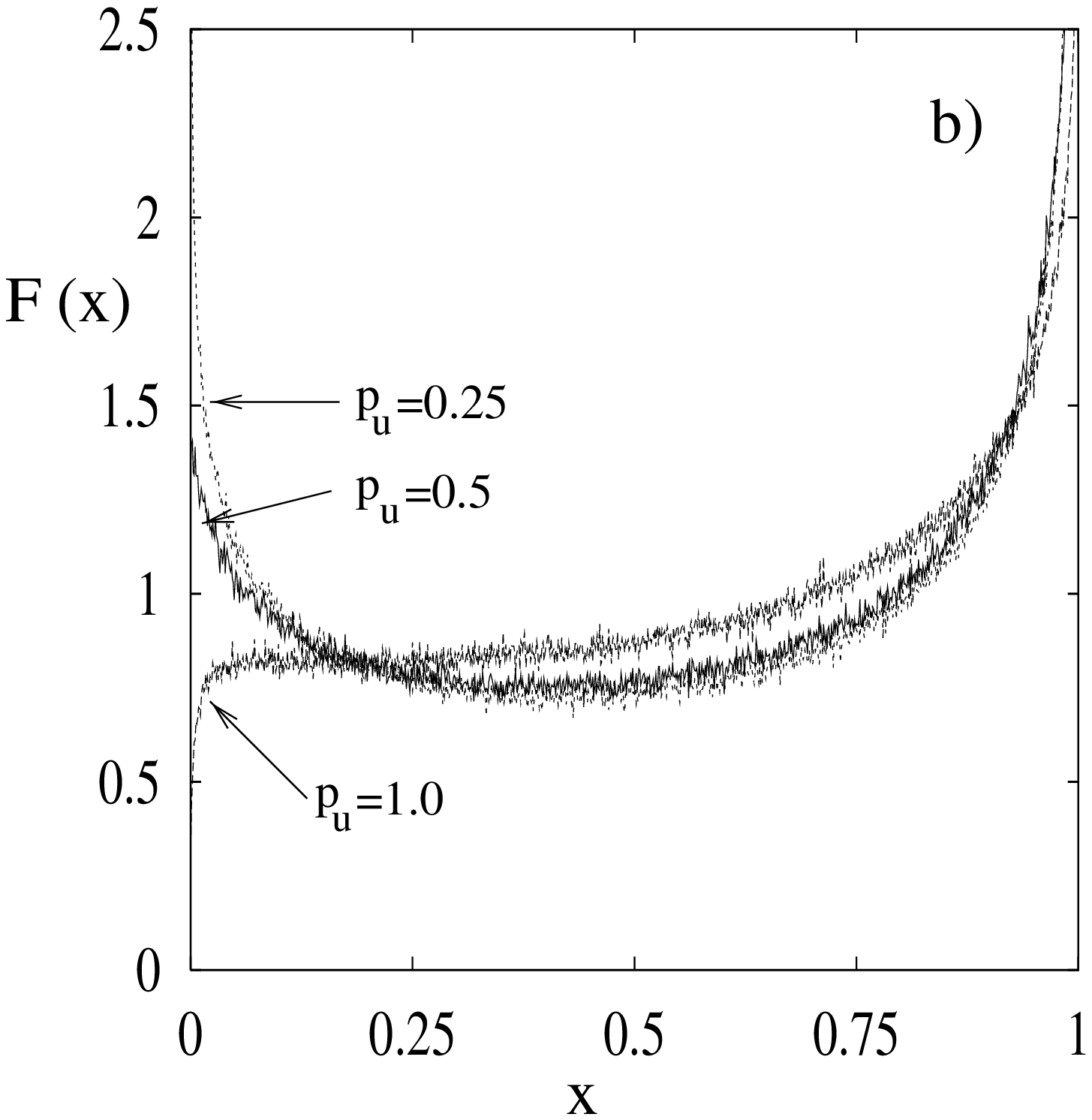}
\end{minipage}
\vspace*{0.5cm}
\caption{a) DST for the $d=1$ KPZ equation 
with white noise
at $t=512$, $t=1024$ and for
bimodal noise at $t=2048$, b)
DST for the SOS DT model with Schwoebel 
barriers at $t=10^3$ 
steps shown for three different values of the parameter
$p_u$ which is the probability for an atom to attach
to a lower step.
The system size was $L=100$, and the averaging 
was made over $2\cdot 10^4$ runs for each curve.
}\label{nonlin}
\end{figure}

We shall now present and briefly discuss the numerically
obtained DST's for two non-linear systems: 1) the one
dimensional KPZ equation and 2)
the Das Sarma-Tamborenea (DT) SOS model with Schwoebel
barriers. Fig.3a shows the KPZ case at different times
and with two different noise types (Gaussian and bimodal),
using the discretization scheme introduced 
in Ref. \cite{bimod}. For the case of Gaussian noise one can see 
that the DST satisfies the general scaling form 
(\ref{scale}) but with 
an asymmetric scaling function $F(x)$
since the $h \to -h$ symmetry is broken, reflecting the
nonlinear character of the KPZ equation. 
The DST for the case of
bimodal noise has a different 
shape to that for Gaussian
noise, which was still evolving for the
largest times we observed ($t \sim 10^{4}$), indicating either
very long crossover times, or else a more complicated 
scaling form.
Fig.3b shows the DST obtained numerically from the 
DT model with Schwoebel barriers \cite{DTB}. 
This system is highly non-linear, exhibits 
mound-formation and coarsening. 
The DST mirrors all these morphological and 
structural characteristics. Nonlinearity 
is obvious from the asymmetric shape. The right end
of the curve has the highest value, meaning that 
the sites are most likely to be found in a positive persistent
state, i.e., they belong to structures that stayed above the
mean height all the time, namely {\em stable mounds}. 
On the contrary, the left end when compared to the right
one, is in minority, showing that 
the {\em stable crevices}, or {\em valleys}
will contain only a small fraction
of the sites, which
points to a mounded morphology with high skewness. 
The fact that a site has a small probability to survive for 
a long time in a crevice, means that the valleys tend 
to disappear during time-evolution, i.e., there must be
{\em coarsening}. This shows the intimate connection 
between the coarsening and 
persistence properties of a interface morphology, which is the 
topic of a separate, forthcoming publication. 

In summary, the DST proves to be very sensitive to 
the details of the morphological dynamics, and  can provide  
crucial information on the non-equilibrium
interface fluctuations. 

The authors are grateful to 
R. Desai, T. Einstein, P.Punyindu,
B. Schmittmann, R.K.P. Zia and E.D. Williams 
for interesting discussions.
Financial support is acknowledged from the Materials
Research Division of the National Science Foundation 
(T.J.N. and Z.T.),  MRSEC (Z.T. and S.D.S.) and 
from the Hungarian Science Foundation, 
T17493 and T19483 (Z.T.).

\end{document}